\documentclass[twocolumn,superscriptaddress,secnumarabic,amssymb, nobibnotes, aps, prl, bm,floatfix,longbibliography]{revtex4-2} 
\usepackage[english]{babel}
\usepackage[utf8]{inputenc}
\usepackage{chemformula} 
\usepackage{graphicx}
\usepackage[colorlinks=true,linkcolor=blue,citecolor=blue]{hyperref}
\usepackage{color}
\usepackage{braket}
\usepackage{physics}
\usepackage{bm}
\renewcommand{\vec}[1]{\mathbf{#1}}
\usepackage{lineno}

\begin{document}

%\title{Optical Manipulation of Real-Space Topological Singularities in the Far-Field Radiation of Photonic Crystals}

%\title{All-optical reconfiguration of far-field singularities in a photonic-crystal laser }
\title{All-optical programming of polarization singularities in a photonic-crystal laser}

\author{Abhishek Padhy}
\affiliation{
Ecole Centrale de Lyon, CNRS, INSA Lyon, Universit\'e Claude Bernard Lyon~1,
CPE Lyon, Institut des Nanotechnologies de Lyon (INL), UMR~5270,
69130 \'Ecully, France
}

\affiliation{
Department of Physics, NISM Institute, University of Namur, Rue de Bruxelles 61, 5000 Namur, Belgium}

\author{Zhiyi Yuan}
\affiliation{
Centre for OptoElectronics and Biophotonics (COEB), School of Electrical and Electronic Engineering, Nanyang Technological University, Singapore 639798
}
\affiliation{
CNRS-International-NTU-Thales Research Alliance (CINTRA), IRL 3288, Singapore }

\author{Mohammed Hamdad}
\affiliation{
Ecole Centrale de Lyon, CNRS, INSA Lyon, Universit\'e Claude Bernard Lyon~1,
CPE Lyon, Institut des Nanotechnologies de Lyon (INL), UMR~5270,
69130 \'Ecully, France }

\author{Panagiotis Nianios}
\affiliation{
Ecole Centrale de Lyon, CNRS, INSA Lyon, Universit\'e Claude Bernard Lyon~1,
CPE Lyon, Institut des Nanotechnologies de Lyon (INL), UMR~5270,
69130 \'Ecully, France
}
\author{Romane Houvenaghel}
\affiliation{
Ecole Centrale de Lyon, CNRS, INSA Lyon, Universit\'e Claude Bernard Lyon~1,
CPE Lyon, Institut des Nanotechnologies de Lyon (INL), UMR~5270,
69130 \'Ecully, France
}
\author{Aziz Benamrouche}
\affiliation{
Ecole Centrale de Lyon, CNRS, INSA Lyon, Universit\'e Claude Bernard Lyon~1,
CPE Lyon, Institut des Nanotechnologies de Lyon (INL), UMR~5270,
69130 \'Ecully, France
}
\author{Nicolas Roy}
\affiliation{
Department of Physics, NISM Institute, University of Namur, Rue de Bruxelles 61, 5000 Namur, Belgium}

\author{Thanh Phong Vo}
\affiliation{
Ecole Centrale de Lyon, CNRS, INSA Lyon, Universit\'e Claude Bernard Lyon~1,
CPE Lyon, Institut des Nanotechnologies de Lyon (INL), UMR~5270,
69130 \'Ecully, France
}
\author{Christian Seassal}
\affiliation{
Ecole Centrale de Lyon, CNRS, INSA Lyon, Universit\'e Claude Bernard Lyon~1,
CPE Lyon, Institut des Nanotechnologies de Lyon (INL), UMR~5270,
69130 \'Ecully, France
}
\author{Xavier Letartre}
\affiliation{
Ecole Centrale de Lyon, CNRS, INSA Lyon, Universit\'e Claude Bernard Lyon~1,
CPE Lyon, Institut des Nanotechnologies de Lyon (INL), UMR~5270,
69130 \'Ecully, France
}
\author{Lotfi Berguiga}
\affiliation{
Ecole Centrale de Lyon, CNRS, INSA Lyon, Universit\'e Claude Bernard Lyon~1,
CPE Lyon, Institut des Nanotechnologies de Lyon (INL), UMR~5270,
69130 \'Ecully, France
}
\author{Michaël Lobet}
\affiliation{
Department of Physics, NISM Institute, University of Namur, Rue de Bruxelles 61, 5000 Namur, Belgium}
\author{S\'egol\`ene Callard}
\affiliation{
Ecole Centrale de Lyon, CNRS, INSA Lyon, Universit\'e Claude Bernard Lyon~1,
CPE Lyon, Institut des Nanotechnologies de Lyon (INL), UMR~5270,
69130 \'Ecully, France
}
\author{Hai Son Nguyen}
\email{hai-son.nguyen@ec-lyon.fr}
\affiliation{
Ecole Centrale de Lyon, CNRS, INSA Lyon, Universit\'e Claude Bernard Lyon~1,
CPE Lyon, Institut des Nanotechnologies de Lyon (INL), UMR~5270,
69130 \'Ecully, France
}
\affiliation{
CNRS-International-NTU-Thales Research Alliance (CINTRA), IRL 3288, Singapore }

\affiliation{
Institut Universitaire de France (IUF), France
}

\date{\today}

\begin{abstract}
Singular optics has emerged as an important research area with diverse applications, yet controlling optical singularities in nanophotonic emitters remains largely constrained by the fixed subwavelength geometry of optical resonators. Here, we circumvent this limitation and demonstrate all-optical programming of real-space polarization singularities in a photonic-crystal laser, while preserving a momentum-space vortex inherited from a symmetry-protected bound state in the continuum. The principle is to use a shaped optical pump to create a smooth mesoscopic potential, whose spatial variations are slow compared with the lattice period. This potential localizes a negative-mass Bloch band into trapped lasing states whose envelope functions, and therefore far-field singularity textures, are defined by the pump geometry. Using a honeycomb photonic crystal supporting a symmetry-protected bound state in the continuum, we achieve room-temperature telecom-band lasing with real-space polarization singularities pinned to the critical points of the envelope function, where its gradient vanishes, and reconfigurable in number and position by pump shaping, while the intrinsic momentum-space singularity at the $\Gamma$ point remains fixed. The experimental observations agree quantitatively with an analytical framework combining the Bloch mode of the photonic crystal with envelope-function theory, establishing optical envelope engineering as a route to programmable structured emission from active photonic lattices.
\end{abstract}


\maketitle

Optical singularities, points where the phase or polarization of light
is undefined, play a central role in structured-light physics and carry
quantized topological charges~\cite{shen_optical_2019, soskin2017singular}. Their control underpins applications in
robust communication \cite{willner2015optical, Ndagano2018}, precision metrology \cite{cheng2025metrology},
and vortex lasing~\cite{hwang2023vortex,mermet-lyaudoz_taming_2023,huang_ultrafast_2020,Gao2024}. In nanophotonics, such singularities may be imposed
externally through phase engineering or arise intrinsically from the
eigenmodes of resonant metasurfaces and photonic crystals (PhCs). Bound
states in the continuum (BICs), whose radiation vanishes by symmetry or
interference, provide a powerful route to generating singular beams
directly from photonic resonances \cite{zhen_topological_2014,mermet-lyaudoz_taming_2023,doeleman_experimental_2018, zhang_observation_2018}.

\begin{figure*}
    \centering
    \includegraphics[width=0.7\linewidth]{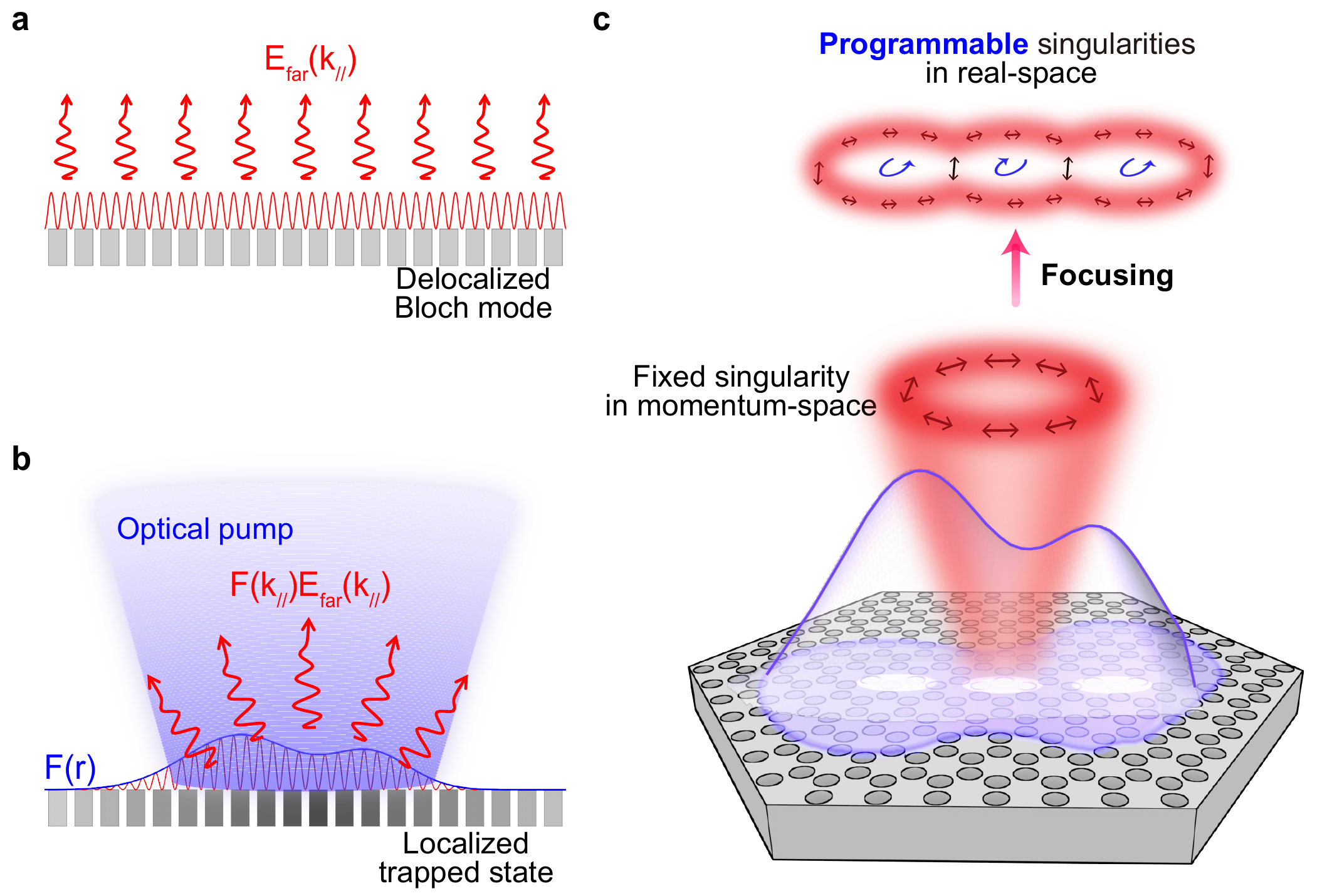}
    \caption{\textbf{Concept:} (\textbf{a}) In the absence of pumping, a Bloch resonance of a PhC slab is delocalized over the lattice and radiates into the continuum with a
fixed momentum-space far-field pattern $\vec E_{\mathrm{far}}(\vec k_\parallel)$.
(\textbf{b}) A shaped optical pump injects carriers and creates a smooth
in-plane potential that localizes the Bloch band into a trapped state with envelope
$F(\vec r_\parallel)$. In momentum space, the trapped-state emission inherits the
Bloch-mode topology but is modulated by the envelope spectrum,
$\vec E^{\mathrm{trap}}_{\mathrm{far}}(\vec k_\parallel)\propto
F(\vec k_\parallel)\vec E_{\mathrm{far}}(\vec k_\parallel)$. (\textbf{c}) Schematic of the all-optical control of the photonic crystal laser with singularities. The far-field singularities in the real-space can be controlled by the optical pump.}
    \label{fig:1}
\end{figure*}

Achieving dynamic control of these singularities remains
challenging because the far-field texture of a Bloch mode is fixed by the unit-cell geometry, and conventional structures offer little reconfigurability.
Thermal tuning in phase-change PhCs allows switching between BIC lasing modes with different topological charges \cite{Tian2022}, and
photo-isomerization in organic PhCs provides a similar mode-switching
mechanism \cite{Yan2025}, but neither approach can reshape the
singularity texture of a given Bloch resonance. Recent optical-control
schemes modulate refractive index at the unit-cell scale: ultrafast
pumping can destroy BIC singularities by breaking in-plane symmetries
\cite{huang_ultrafast_2020}, and gain–loss perturbations suppress vortex
emission \cite{Gao2024,Wu2024}. Sub-unit-cell pumping can even alter the
response of individual elements \cite{Aigner2025}, but remains limited to
passive linewidth tuning and does not enable continuous control of
far-field singularities. Fundamentally, all these schemes attempt to
reconfigure the unit cell, an operation constrained by the diffraction
limit.
In this work, we introduce an alternative approach, instead of
modifying the subwavelength Bloch function, we reconfigure the
mesoscopic envelope of a PhC resonance using optical pumping (see Fig.~\ref{fig:1}). Carrier injection induced by the pump generates a smooth in-plane potential, which confines a Bloch band into localized states whose envelope functions follow a two-dimensional Schrödinger equation. By tailoring the pump profile, the induced potential is varied, enabling arbitrary control over the envelope function. Since the emitted field is given by the product of the Bloch mode’s far field and its envelope, this approach allows reconfigurable far-field singularities without modifying the underlying photonic crystal structure.

As a proof of concept, we use a honeycomb PhC supporting the lowest-energy
band with isotropic negative-mass dispersion and a monopolar BIC at the $\Gamma$-point 
($\vec{k}_\parallel$= $\vec{0}$). 
Optical pumping forms tunable potentials that trap this band and yield lasing trapped states whose real-space far-field polarization
singularities are pinned at the critical points of the envelope function and can be programmed by the pump geometry, whereas the momentum-space singularity remains fixed. Lasing occurs at room temperature and telecom wavelengths, and the experimental measurements agree quantitatively with envelope-function theory, establishing a general route to reconfigure singular beams in active photonic lattices.

\section{Results}
\subsection{Theoretical framework}
We begin by considering a generic guided resonances of a PhC slab in the absence of optical pumping.  
For a fixed in-plane wave vector $\vec{k}_\parallel = (k_x,k_y)$, the corresponding Bloch mode inside the slab can be expanded over a finite basis  consisting of guided modes $\{\ket{n}\}$ of an effective homogeneous slab, each carrying a reciprocal lattice vector $\vec{G}_n$. The guided modes in the basis share the same out-of-plane profile $u(z)$ and differ only in their in-plane Bloch harmonics. Factoring out the global Bloch phase $e^{i\vec{k}_\parallel\cdot\vec{r}_\parallel}$ for compactness, the near field can be written as $\vec{E}_{\text{near},\vec{k}_\parallel}(\vec{r}_\parallel,z)= u(z)\sum_{n} A_n(\vec{k}_\parallel)\, e^{i\vec{G}_n\cdot\vec{r}_\parallel}\,
    \vec{p}_n$, where $\vec{r}_\parallel=(x,y)$ denotes the in-plane coordinate, $\{\vec{p}_n\}$ are the polarization vectors of the guided-mode basis. Here,  $A_n(\vec{k}_\parallel)$ are the expansion coefficients obtained by diagonalizing the effective non-Hermitian Hamiltonian of the PhC slab~\cite{nguyen_generalized_2025}. Radiation into the continuum is obtained by projecting this Bloch mode onto the outgoing Fabry--Pérot channel of the unpatterned slab [Fig.~\ref{fig:1}(a)]. In the single-channel regime relevant here (subwavelength lattice, thus the zeroth diffraction order inside the light cone is the only leaky channel), the radiated field in direction $\vec{k}_\parallel$ is proportional to the same guided-mode coefficients, but without the fast lattice harmonics $e^{i\vec{G}_n\cdot\vec{r}_\parallel}$~\cite{nguyen_generalized_2025}: $\vec{E}_{\text{far}}(\vec{k}_\parallel)\;\propto\;\sum_{n}\sqrt{\gamma_n}A_n(\vec{k}_\parallel)\,\vec{p}_n$
where $\gamma_n$ is the radiative loss rate of the guided mode $\ket{n}$ when folded inside the light cone and coupled to the radiative continuum. The polarization texture and the existence of BICs are encoded in this farfield vector, and the radiative rate $\gamma(\vec{k}_\parallel)$ of the resonance is given by the emitted intensity $\vert\vec{E}_{\text{far}}\vert^2$. 
\begin{figure}[ht!]
    \centering
    \includegraphics[width=0.9\linewidth]{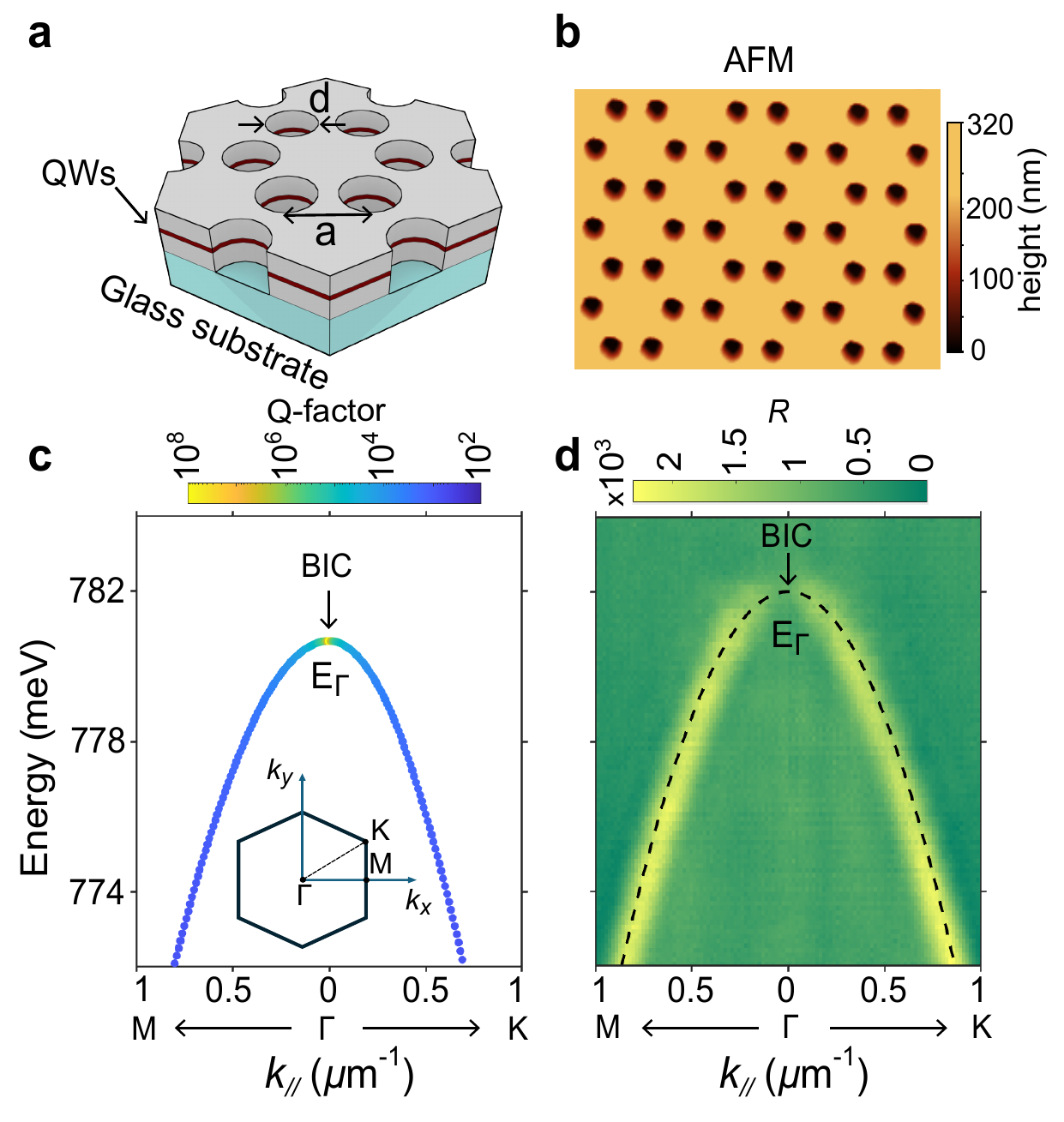}
\caption{\textbf{The fabricated photonic crystal device and its dispersion:} \textbf{(a)} Schematic of the fabricated honeycomb PhC slab with quantum wells (QWs) integrated.  
The lattice has a period \(\Lambda = a\sqrt{3}\) with lattice constant \(a = 445~\mathrm{nm}\), hole diameter \(d = 335~\mathrm{nm}\), and total membrane thickness \(t = 240~\mathrm{nm}\).
\textbf{(b)} Atomic-force-microscopy (AFM) image of the fabricated surface.
\textbf{(c)} Numerically computed band structure using Legume, showing the lowest-energy band that hosts a monopolar symmetry-protected BIC at the \(\Gamma\)-point.  
The band is nearly isotropic and displays a negative effective mass.
\textbf{(d)} Experimental reflectivity measurements along the \(\Gamma\!-\!\text{K}\) and \(\Gamma\!-\!\text{M}\) directions, exhibiting the vanishing radiative loss at \(\Gamma\) which is characteristic of the BIC.  
The dashed line represents a parabolic fit to the measured dispersion.
}
\label{fig:2}
\end{figure}
 \begin{figure*}[ht!]
    \centering
    \includegraphics[width=1.0\linewidth]{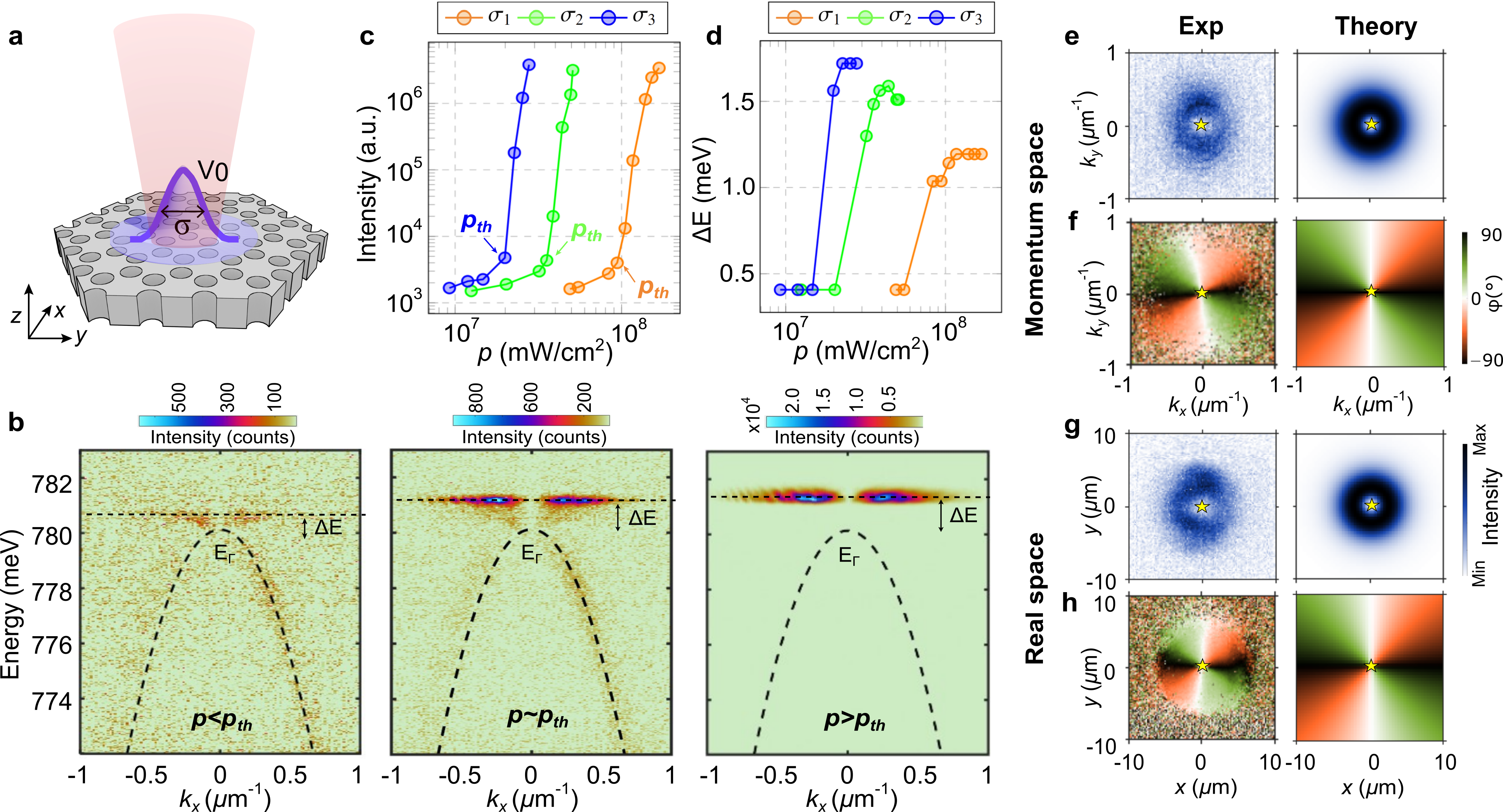}
    \caption{\textbf{Lasing characterization and farfield measurements for single pump spot:} \textbf{(a)} Schematic of single spot pumping on the PhC with spot size $\sigma$. The pump creates a potential $V(\vec{r}_\parallel)$, which induces a trapped-state envelope function $F(\vec{r}_\parallel)$. \textbf{(b)} Experimental energy-momentum dispersion along $k_x$ below ($p<p_{th}$), around ($p \sim p_{th}$) and above ($p>p_{th}$) threshold pump power density ($p_{th}$) for spot size $\sigma_1= 3.2~\mu$m . The fundamental mode is fitted with a parabola with $\alpha = -13.4$~meV·$\mu$m$^2$, and the confinement energy ($\Delta$E) is the energy difference between the trapped state and parabolic monopolar band at $\Gamma$-point (E$_{\Gamma}$).  \textbf{(c)} Total integrated trapped-state intensity vs pump power density $p$ for three pump spot size $\sigma_1= 3.2$ $\mu$m, $\sigma_2= 7.1$ $\mu$m, $\sigma_3= 11.1$ $\mu$m. The threshold pump power density $p_{th}$ for the given spot sizes are $p_{th}(\sigma_1)= 9.4\cdot10^{7}$ mW/cm$^{2}$, $p_{th}(\sigma_2)= 3.6\cdot10^{7}$ mW/cm$^{2}$ and 
    $p_{th}(\sigma_3)= 2\cdot10^{7}$ mW/cm$^{2}$ . \textbf{(d)} Confinement energy $\Delta$E vs pump power for the three spot size. \textbf{(e–h)} Far-field characterization of the trapped state above threshold for $\sigma=3.2~    \mu$m. Panels show (left) experimental momentum-space and real-space intensity and polarization textures and (right) corresponding analytical/theoretical predictions from the envelope-function model. The polarization texture maps show the polarization orientation $\phi$ and the yellow stars mark the polarization singularities with topological charge $q=1$. The theoretical results were calculated with the parameter $V_0=$ 4.8 meV. Refer to Ext. Fig.~\ref{fig:Extended_Figure_1} for the polarization resolved farfield lasing images  and Ext. Fig.~\ref{fig:EF-Sim-Vortex} for farfield results computed via FDTD.} 
    \label{fig:3}
\end{figure*}
The complex eigenfrequency of the guided resonance can be described as $\hbar\omega(\vec{k}_\parallel) \simeq \hbar\omega_0 + \frac{\hbar |\vec{k}_\parallel|^2}{2m}- i\,\hbar\gamma(\vec{k}_\parallel)$,
where $m$ is the effective mass of the Bloch resonance. 
In general, the photonic band can exhibit anisotropy, but here, the guided resonances have been chosen such that the effective mass $m$ is negative and isotropic. Under optical pumping, photo-generated carriers produce a carrier-induced optical nonlinearity that leads to a spatially varying refractive-index change $\Delta n(\vec{r}_\parallel)<0$ \cite{Bennett1990}. This smooth index decrease maps onto an effective in-plane blueshift potential $V(\vec{r}_{\parallel})$ for the Bloch band \cite{Leonard2002,Fushman2007}, thereby trapping the lowest-energy states in a pump-defined landscape.
Within the effective-mass approximation, considering a smooth varying potential $V(\vec{r}_{\parallel})$  and neglecting losses in the in-plane eigenproblem for  small $\gamma(\vec{k}_\parallel)$, the in-plane dynamics are governed by a standard Hermitian Schr\"odinger equation for the envelope function $F(\vec{r}_\parallel)$\cite{Ihn2009}:
    $-\frac{\hbar^2}{2m}\,\nabla_{\vec{r}_\parallel}^2 F(\vec{r}_\parallel)
    + V(\vec{r}_\parallel)\,F(\vec{r}_\parallel)
    = \Delta E\, F(\vec{r}_\parallel).$
Because $m<0$, a positive potential $V(\vec{r}_\parallel)$ has a confining effect and localizes light under the pump spot, forming trapped states. 
Each trapped state corresponds to a solution $F(\vec{r}_\parallel)$ of the Schrödinger equation given above; its Fourier transform $F(\vec{k}_\parallel)=\text{FT}[F(\vec{r}_\parallel)]$ specifies how the Bloch resonances are superposed in momentum space.
The near field of a trapped state can thus be written as a superposition of Bloch modes weighted by $F(\vec{k}_\parallel)$:
\begin{equation}
    \vec{E}^{\text{trap}}_{\text{near}}(\vec{r}_\parallel)
    =
    \sum_{\vec{k}_\parallel}
    F(\vec{k}_\parallel)\,
   \vec{E}_{\text{near},\vec{k}_\parallel}(\vec{r}_\parallel,z),
    \label{eq:nearfield_trap}
\end{equation}
or, in the continuum limit, as an integral over $\vec{k}_\parallel$. The corresponding far field of the trapped state in momentum space is
\begin{equation}
    \vec{E}^{\text{trap}}_{\text{far}}(\vec{k}_\parallel)
    =
    F(\vec{k}_\parallel)\,\vec{E}_{\text{far}}(\vec{k}_\parallel),
    \label{eq:farfield_trap_k}
\end{equation}
that is, the original Bloch far-field pattern is modulated by the envelope spectrum $F(\vec{k}_\parallel)$ [see Fig.~\ref{fig:1}(b)]. Finally, the far field of the trapped state in real space is obtained by the inverse Fourier transform of Eq.~\eqref{eq:farfield_trap_k} as follows:
\begin{equation}
    \vec{E}^{\text{trap}}_{\text{far}}(\vec{r}_\parallel)
    =
    \text{FT}^{-1}
    \bigl[
        F(\vec{k}_\parallel)\,\vec{E}_{\text{far}}(\vec{k}_\parallel)
    \bigr].
    \label{eq:farfield_trap_rho}
\end{equation}
Equations~\eqref{eq:nearfield_trap}--\eqref{eq:farfield_trap_rho} make the central mechanism explicit: the singularities of the radiated field are dictated jointly by the Bloch resonance, through 
$\vec{E}_{\text{far}}(\vec{k}_\parallel)$, and by the pump-induced envelope, through $F(\vec{k}_\parallel)$.  
The Bloch contribution determines the intrinsic momentum-space topology carried by the underlying band, whereas the envelope reshapes the distribution of momenta and thereby controls the number, position, and structure 
of the singularities in the real space.

A special case arises when the underlying Bloch resonance is a symmetry-protected BIC. Since such modes satisfy $\vec{E}_{\mathrm{far}}(\vec{k}_\parallel=\vec{0})=\vec{0}$, Eq.~\eqref{eq:farfield_trap_k} ensures that any trapped state constructed from them also obeys $\vec{E}^{\mathrm{trap}}_{\mathrm{far}}(\vec{k}_\parallel=\vec{0})=\vec{0}$, independently of the envelope function. The momentum-space singularity at the $\Gamma$ point is therefore preserved under arbitrary optical pumping. In contrast, the real-space far field directly reflects the envelope of the trapped state. For symmetry-protected BICs of winding number $+1$, commonly encountered in photonic crystal slabs with $C_4$ or $C_6$ symmetry, the Bloch radiation scales as $\vec{E}_{\mathrm{far}}(\vec{k}_\parallel)\propto \vert\vec{k}_\parallel\vert\,\mathbf{u}_\varphi$. Consequently, the radiation spectrum of a trapped state with envelope $F(\vec{k}_\parallel)$ becomes $\vert\vec{k}_\parallel\vert\,F(\vec{k}_\parallel)\,\mathbf{u}_\varphi$. According to Eq.~\eqref{eq:farfield_trap_rho}, the corresponding real-space far field is therefore given by the inverse Fourier transform, leading to:
\begin{equation}
\vec{E}^{\text{trap}}_{\text{far}}(\vec{r}_\parallel)
\propto
\hat{\mathbf z}\times\nabla F(\vec{r}_\parallel).
\label{eq:monopolar_farfield_realspace}
\end{equation}

Equation~\eqref{eq:monopolar_farfield_realspace} shows that the real-space far field is governed entirely by the spatial variation of the envelope function $F(\vec{r}_\parallel)$. In particular, polarization singularities of $\vec{E}^{\text{trap}}_{\text{far}}(\vec{r}_\parallel)$ occur at points where $\nabla F(\vec{r}_\parallel)=0$, corresponding to critical points (maxima, minima, or saddle points) of the envelope. These results reveal a clear separation between momentum-space and real-space topologies: the Bloch resonance fixes the polarization vortex at $\Gamma$ in $\vec{k}_\parallel$-space, whereas the pump-defined envelope determines the number, position, and spatial distribution of polarization singularities in real space. 

\subsection{Proof of Concept}
%\emph{Proof of concept.| }
To demonstrate envelope-function control experimentally, we use a
two-dimensional honeycomb photonic crystal slab patterned into an
InP/InAsP/InP multiple–quantum-well membrane [Fig.~\ref{fig:2}(a–b)].  
The heterostructure is grown by molecular beam epitaxy, the active
InP-based membrane is molecularly bonded onto a glass substrate, and the
PhC is defined by electron-beam lithography followed by reactive-ion
etching.

The band-structure of the PhC slab was obtained numerically using the guided-mode-expansion solver Legume \cite{Zanotti2024} and experimentally via angle-resolved photoluminescence measurements [Fig.~\ref{fig:2}(c-d)](see Methods for setup details). They show that the lowest-energy (monopolar) band of the fabricated honeycomb PhC is nearly isotropic and follows a parabolic dispersion with negative effective mass. At the $\Gamma$-point, this band hosts a symmetry-protected monopolar BIC, evidenced by the vanishing radiative loss
$\gamma(\vec{k}_\parallel)\propto|\vec{k}_\parallel|^{2}$.
%[Fig.~\ref{fig:2}(c–d)].
A full analytical description of this Bloch resonance, including the effective non-Hermitian Hamiltonian and its eigenmodes, is provided in Methods. In the following, we use this monopolar band to engineer the trapped states, whose envelope function is entirely shaped by the optical pump. However, we note that Equations~(\ref{eq:farfield_trap_k})–(\ref{eq:farfield_trap_rho}), which factorize the trapped-state radiation as
$\vec E^{\mathrm{trap}}_{\mathrm{far}}(\vec k_\parallel)=F(\vec k_\parallel)\vec E_{\mathrm{far}}(\vec k_\parallel)$,
are quantitatively accurate only when the target Bloch band is well isolated from other leaky bands over the momentum range sampled by $F(\vec k_\parallel)$, so that interband mixing can be neglected.
This condition is well satisfied for the monopolar band studied here: it is spectrally isolated around $\Gamma$, and its closest neighbor is the hexapolar band which also exhibits a BIC at $\Gamma$.

 \begin{figure}[ht!]
    \centering
    \includegraphics[width=1.0\linewidth]{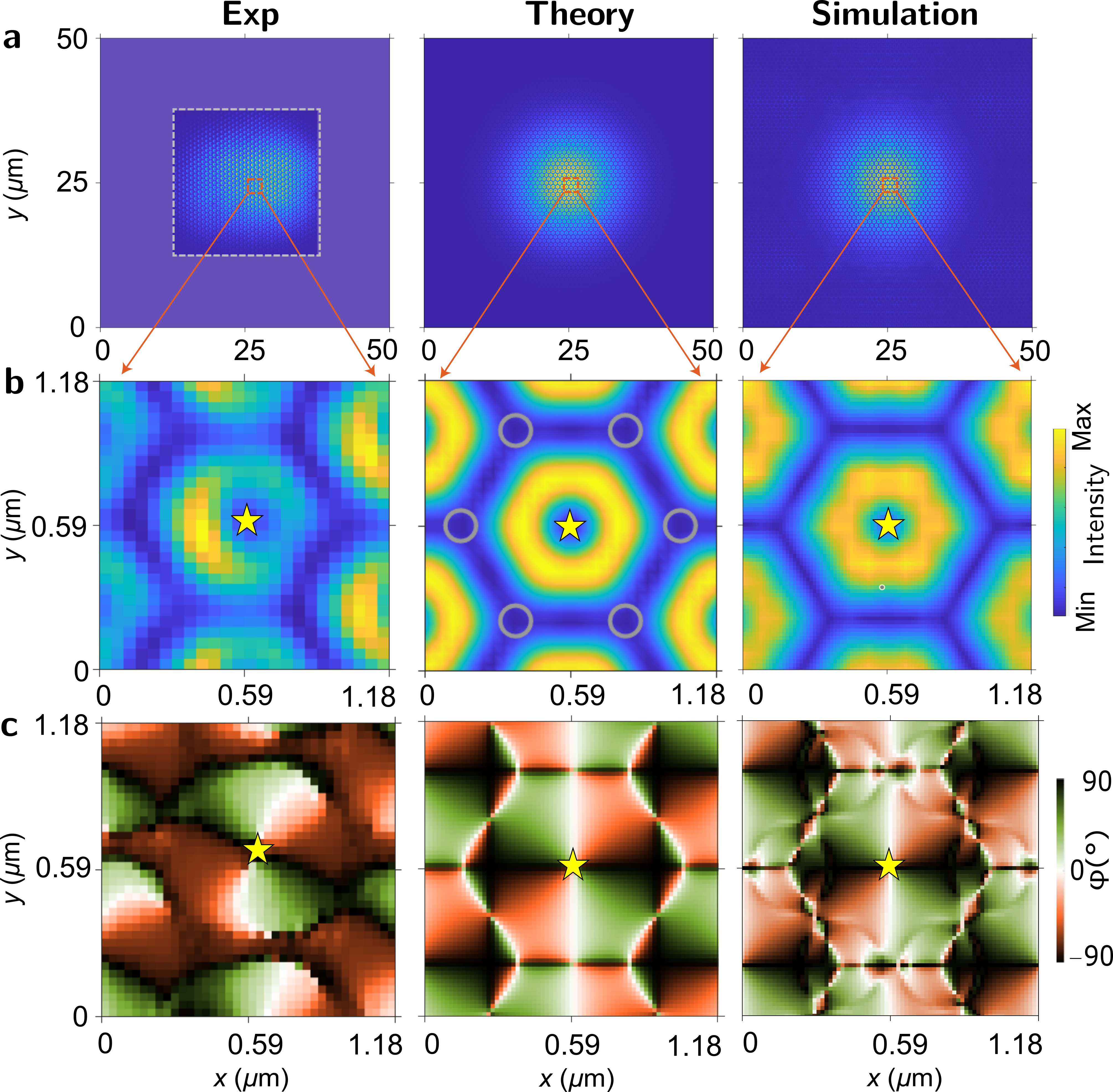}
    \caption{\textbf{Nearfield maps for single pump:} \textbf{(a)} Large-area SNOM (left panel), analytical model (middle), and FDTD numerical (right) of near-field map of the trapped state under a single Gaussian pump. The white dashed line box represents the measurement region.   
\textbf{(b)} Zoomed-in near-field amplitude showing the fast spatial
oscillations of the Bloch resonance on the scale of the photonic crystal
lattice.  
\textbf{(c)} Polarization-resolved mapping of the in-plane electric field.}
    \label{fig:4}
\end{figure}
The sample was excited using a femtosecond pulsed laser at 800 nm and room temperature, shaped by a spatial light modulator (SLM). The pump was focused on a single Gaussian spot on the PhC slab [Fig.~\ref{fig:3}(a)], injecting carriers into the InP/InAsP/InP multiple–quantum-well membrane and creating a smooth negative variation of refractive-index that acts as a confining potential for the lowest-energy Bloch band. This potential can be modeled as
$V(\vec r_\parallel)=V_0 e^{-|\vec r_\parallel|^2/\sigma^2}$, where the depth $V_0$ is set by the injected carrier density, and the trap size $\sigma$ is controlled precisely by the SLM.

Figure~\ref{fig:3}(b) shows the measured dispersion with increasing pump power density. At low pump power density, the emission follows the unperturbed band with inverted parabolic dispersion $\text{E}(k_x)=\text{E}_\Gamma+\alpha k_x^2$; a fit gives $\alpha=-13.4\,\mathrm{meV\,\mu m^2}$ where E$_{\Gamma}=$ 780.125 meV, confirming the negative effective mass. Upon increasing the pump power, a dispersionless line appears above E$_\Gamma$, indicating the formation of a trapped state with confinement energy $\Delta$E. Below threshold, the trapped mode blueshift increases with pump power as injected carriers deepen the effective confining potential [Fig.~\ref{fig:3}(c)]. Above threshold, the blueshift saturates as stimulated emission clamps the carrier density. Additional pump power, therefore, produces photons rather than increasing the carrier reservoir, fixing the confinement depth at $V_o\approx 4.8 $ meV (see Methods) .

The total flatband intensity showed a characteristic $S$-shaped dependence on the pump power density, demonstrating single-mode lasing. Notably, the lasing threshold decreases when the pump-spot size increases [$\sigma_1<\sigma_2<\sigma_3$ in Fig.~\ref{fig:3}(c)]. A smaller excitation spot produces a broader momentum-space envelope, leading to  stronger overlap with the high-$\text{k}$ components of the Bloch mode, whose radiative losses scale as $|\vec{k}_\parallel|^{2}$. The trapped-state radiative loss therefore scales approximately as $1/\sigma^{2}$. Consequently, smaller traps require a higher carrier density at threshold, whereas larger traps exhibit weaker loss and lase at lower gain. The saturated confinement energy $\Delta$E above the threshold also increases with the pump-spot diameter [Fig.~\ref{fig:3}(d)]. Indeed, a larger $\sigma$ creates a broader and shallower potential, and in the negative-mass Schrödinger model, the trapped-state energy shifts upward.

We investigate the far-field radiation pattern for $\sigma=$ 3.2~$\mu$m  by examining the corresponding far-field intensity and polarization textures in both momentum and real space above threshold. The measurements, together with the analytical predictions of the envelope-function model, are shown in Fig.~\ref{fig:3}(e–h).The radiation patterns remain unchanged for other spot sizes. The momentum-space lasing emission exhibits a polarization vortex centered at $\vec{k}_\parallel=\vec{0}$ [Fig.~\ref{fig:3}(e–f)], characteristic of the monopolar BIC of the underlying Bloch mode \cite{nguyen_generalized_2025}. As follows from Eq.~(\ref{eq:farfield_trap_k}), the radiative amplitude vanishes at the $\Gamma$-point irrespective of the envelope function. The momentum-space singularity is therefore preserved independently of the pump profile. In real space, however, the far-field pattern is set by the envelope of the trapped state. According to Eq.~(\ref{eq:monopolar_farfield_realspace}), the emitted field is proportional to the spatial gradient of the envelope function. For a radially symmetric envelope $F(\vec{r}_\parallel)=F(|\vec{r}_\parallel|)$ produced by a symmetric pump spot, the far field reduces to $F'(|\vec{r}_\parallel|)\,\vec{u}_\varphi$. In the case of the Gaussian pump used here, this results in a donut-shaped real-space intensity distribution with a single polarization singularity at the center, in agreement with the measurements shown in Fig.~\ref{fig:3}(g–h).
%\subsection{Near-field measurements}

Near-field imaging with scanning near-field optical microscopy (SNOM) \cite{Vo_2010,Vo_2012}
provides direct access to both the mesoscopic envelope and the microscopic Bloch component of the trapped state.
The large-area SNOM map in Fig.~\ref{fig:4}(a) reveals a Gaussian-like intensity profile localized beneath the pump,
with a maximum at the pump center and a smooth decay, in agreement with the confined state predicted by the
effective-mass model. The slight elongation of the envelope is likely caused by a weak ellipticity of the excitation
focus in the SNOM setup. Zoomed-in views [Figs.~\ref{fig:4}(b–c)] resolve the rapid subwavelength oscillations
associated with the Bloch mode. Within each unit cell, the measured near field displays the characteristic azimuthal
rotation expected for the magnetic-monopole resonance: the in-plane field winds around the hexagon center with a full
$2\pi$ polarization rotation. Polarization-resolved SNOM measurements [Fig.~\ref{fig:4}(c)] further corroborate this picture,
showing a clear vortex-like winding of the local electric-field vector. Together, these results demonstrate that the trapped state preserves the microscopic polarization texture of the
monopolar Bloch resonance while acquiring a mesoscopic envelope set by the optically induced potential.
The SNOM observations are in excellent agreement with both the analytical envelope-function model and full-wave
finite-difference time-domain (FDTD) simulations. In the FDTD calculations, we emulate the optically induced potential
by introducing a Gaussian refractive-index perturbation,
$n(\vec r_\parallel)=n_0+\delta_n \exp(-|\vec r_\parallel|^2/\sigma^2)$, with $n_0=3.162$ and $\delta_n=-0.022$.

\begin{figure*}
    \centering
    \includegraphics[width=1\linewidth]{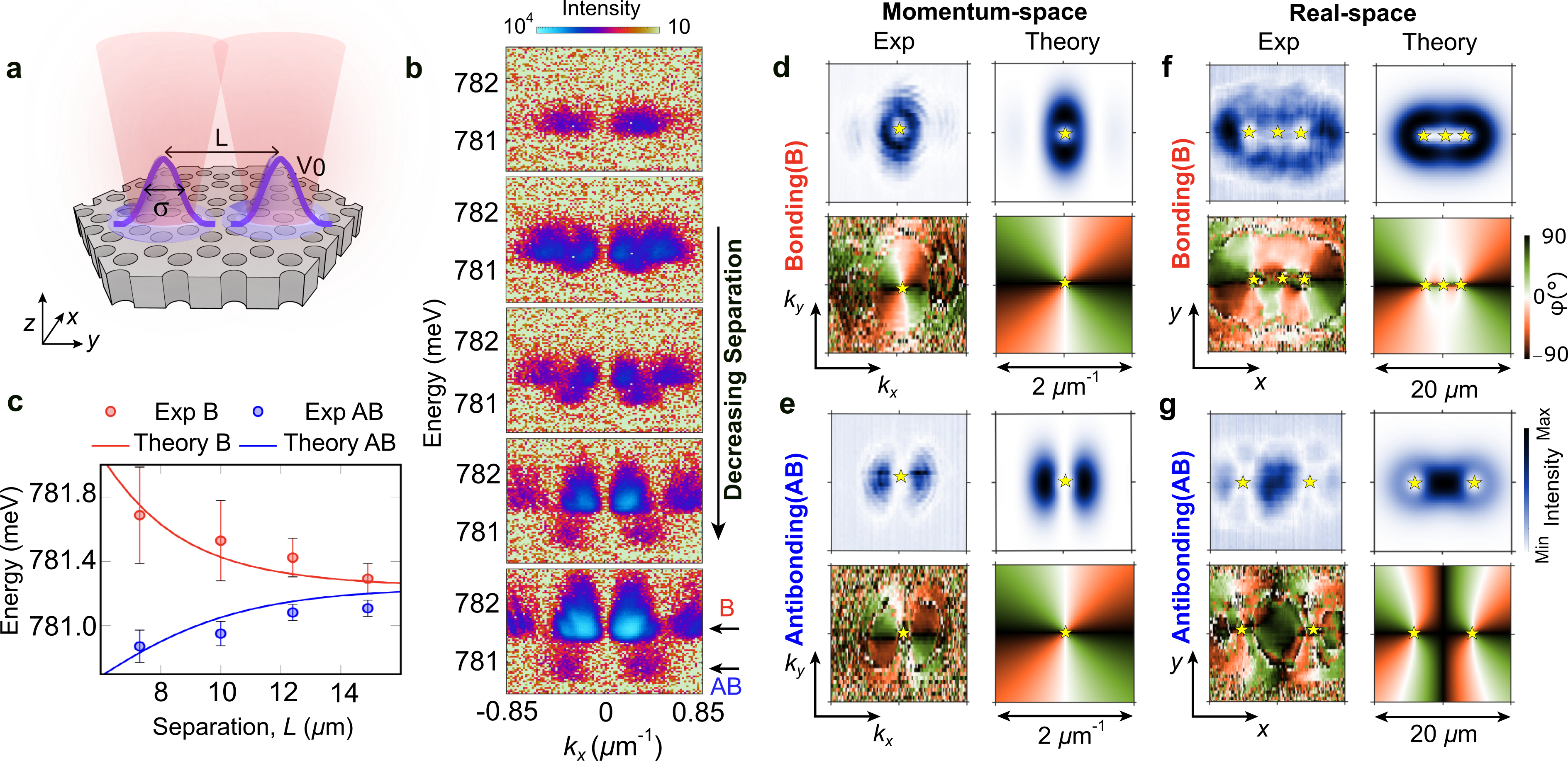}
\caption{\textbf{Two-spot pumping and reconfigurable far-field singularities.}
\textbf{(a)} Schematic illustrating two trapped states induced by a double pump configuration with variable separation distance $L$ and fixed spot waist $\sigma = 3.25~\mu$m.
\textbf{(b)} Energy–momentum dispersion along $k_x$ for different values of $L$ (16.5, 14.9, 12.4, 10, and 7.2~$\mu$m), shown from top to bottom in decreasing order. At the smallest separation ($L = 7.2~\mu$m), two distinct modes emerge, identified as the antibonding (AB) and bonding (B) states.
\textbf{(c)} Evolution of the AB and B mode energies as a function of separation $L$. Experimental data (points with error bars) are compared with theoretical predictions (solid lines).
Far-field intensity distributions in momentum and real space for $L = 7.2~\mu$m are shown for the bonding mode \textbf{(d,f)} and antibonding mode \textbf{(e,g)}. Experimental results (left panels) are in good agreement with analytical/theoretical calculations (right panels). The polarization texture maps show the polarization orientation $\phi$ and the yellow stars mark the polarization singularities with topological charge $q=1$. Theoretical results are obtained using $V_0 = 4.6$ meV. See Ext. Fig.~\ref{fig:Extended_Figure_2} for polarization-resolved far-field lasing images, Ext. Fig.~\ref{fig:EF-Sim-Vortex} for FDTD-based simulations and Ext. Fig.~\ref{fig:EF-Near-Field-couplemode} for nearfield maps via FDTD and analytical model.
}
    \label{fig:5}
\end{figure*}

\subsection{Reconfigurable singularities}

As indicated by Eq.~(\ref{eq:monopolar_farfield_realspace}), polarization singularities arise at spatial locations where $\nabla F(\vec r_\parallel)=0$, that is, at the critical points of the envelope function. This relation provides a direct pathway for dynamic tailoring the spatial distribution of singularities by shaping the pump-defined envelope. To illustrate this principle, we patterned the pump into two Gaussian spots separated by a distance $L$, forming a diatomic photonic “molecule’’ [Fig.~\ref{fig:5}(a)]. Each pump spot defines an optically induced trap that plays the role of an “atom,” while their overlap enables tunnel-like coupling controlled by $L$. The pump waist was fixed to $\sigma=3.25\,\mu$m, and the device was driven above threshold into the lasing regime.

The measured energy–momentum dispersions for decreasing $L$ are shown in
Fig.~\ref{fig:5}(b). For large separations (top row), only a single trapped-state
branch is visible, corresponding to two nearly independent localized
states. As $L$ decreases, coupling between the traps increases and the
lasing mode splits into antibonding (AB) and bonding (B) branches,
 analogous to the molecular eigenstates of a double-well potential~\cite{AtkinsFriedman_MQM_5ed}. The
dependence of their energies on $L$ is plotted in Fig.~\ref{fig:5}(c)
and agrees quantitatively with the effective-mass Schrödinger model for
a two-well potential
$V(x,y)=V_0\!\left[e^{-((x-L/2)^2+y^2)/\sigma^2}+e^{-((x+L/2)^2+y^2)/\sigma^2}\right]$,
using $V_0$ obtained from the method discussed for a single spot. This establishes that
the lasing branches correspond to the B and AB envelope eigenstates of
pump-induced potential landscape.

Because the underlying Bloch resonance is a monopolar BIC with
$\vec{E}_{\mathrm{far}}(\vec{0})=\vec{0}$, both B and AB modes retain a
single polarization vortex pinned at the $\Gamma$ point, as seen in the
momentum-space measurements of Fig.~\ref{fig:5}(d--e). In real space,
however, the two modes exhibit different polarization-singularity
structures. This difference follows from the critical points of their
two-dimensional envelope functions (see Supplemental Information for detailed discussion). For the antibonding mode, the envelope changes sign between the two pump-defined traps and exhibits
two critical points in the measured plane, associated with the two
localized lobes. In the present geometry, these correspond to one
maximum and one saddle point. For the bonding mode, the envelope
exhibits three critical points: two maxima localized near the two pump
spots and one saddle point at the centre of the photonic molecule.
According to Eq.~(\ref{eq:monopolar_farfield_realspace}), these critical
points satisfy $\nabla F(\vec r_\parallel)=0$ and therefore appear as
polarization singularities in the emitted real-space field. Consistently,
the AB mode displays two singularities, whereas the B mode exhibits
three singularities [Fig.~\ref{fig:5}(f--g)]. The analytical far-field patterns evaluated from the envelope-function
model [right column of Fig.~\ref{fig:5}(d–g)] reproduce the measured
vortex configurations with no adjustable parameters. These results
demonstrate that while the monopolar BIC fixes the topology in
momentum space, the pump-defined envelope determines the spatial
distribution of polarization singularities in real space.

To demonstrate the scalability of envelope-function control, we also
investigated a configuration using three identical pump spots arranged
linearly with equal separation $L = 7.5~\mu\mathrm{m}$.  
The three pump spots generate a composite potential consisting of three
overlapping Gaussian wells, which supports three trapped eigenstates—the
photonic analogue of the three molecular orbitals in a linear
triatomic chain.  
These states appear as three distinct lasing modes and correspond to the
three lowest solutions of the effective-mass Schr\"odinger equation for a triple-well potential. Extended Figure~\ref{fig:Extended_Figure_3} compares the experimental and theoretical far-field
emission patterns of the three trapped modes in both real space and
momentum space, together with the associated polarization textures
obtained from polarization-resolved tomography. As expected, all three modes preserve the same momentum-space vortex at
$\Gamma$, inherited from the monopolar BIC of the underlying Bloch
resonance. In contrast, each trapped mode exhibits a distinct real-space polarization-singularity
pattern determined by the nodal structure of its envelope function:
the ground state (Mode~1) displays five singularities, the first excited
state (Mode~2) displays two singularities, and the second excited state
(Mode~3) displays three singularities. The experimental maps and analytical predictions show excellent
quantitative agreement in both intensity and polarization textures,
including the singularity locations indicated by the yellow stars in Extended Fig.~\ref{fig:Extended_Figure_3}.

These results demonstrate that envelope-function engineering enables reconfigurable real-space singularities—including their number, type, and position—while the momentum-space topology remains protected by the symmetry of the underlying Bloch mode. Adjusting the pump separation $L$ continuously tunes the molecular hybridization and therefore the far-field vortex pattern. More broadly, because the effective potential is written optically, the approach naturally extends to more complex pump geometries (e.g., multi-spot arrays, asymmetric patterns, and dynamically modulated profiles), providing additional degrees of freedom to sculpt the trapped-state landscape and the resulting singularity textures on demand. This concept thus offers a scalable route to programmable singular-beam generation in active photonic crystals.

\section{Discussion}

We demonstrated an all-optical mechanism for reconfiguring far-field singularities in active photonic crystals by tailoring the mesoscopic envelope of a Bloch resonance. This approach bypasses the need to modify subwavelength unit-cell geometries and instead uses pump-induced carrier potentials to localize a Bloch band into trapped states described by a two-dimensional Schrödinger equation. In a honeycomb PhC supporting a monopolar BIC, we demonstrated that the intrinsic momentum-space singularity remains fixed, whereas the real-space polarization texture can be continuously reshaped through the envelope function. Multi-spot pumping enables trapped states hosting controllable numbers and positions of singularities, which is in quantitative agreement with the analytical theory. Because the mechanism relies solely on carrier injection, reconfiguration can, in principle, operate on picosecond timescales \cite{Leonard2002,Fushman2007,Yu2014,Shcherbakov2017}. Extensions toward dynamic pump modulation and programmable multi-spot pumping could enable reconfigurable tight-binding Hamiltonians for quantum simulation~\cite{AspuruGuzik2012,Grass2025} and laser-array architectures with controllable coupling for neuromorphic computing~\cite{Chen2023,Ji2025}, further expanding the possibilities of ultrafast programmable structured-light emitters.

\section{ACKNOWLEDGEMENTS}
We thank Philippe Regreny for growing epitaxial structures. Wafer bonding technology was supplied by CEA-LETI.  This work was supported by the French National Research Agency (ANR) under the projects NANOEC (ANR07-NANO-036), POLAROID (ANR-24-CE24-7616-01), and SUPER-HERO (ANR-25-CE24-4066). A.P. was supported by the LABEX iMUST of the University of Lyon (ANR-10-LABX-0064) through the “Stage d’Innovation 2023–24” program. M.L. is a Research Associates of the Fonds de la Recherche Scientifique—FNRS. Computational resources have been provided by the Consortium des Equipements de Calcul Intensif (CECI), ´ funded by the Fonds de la Recherche Scientifique de Belgique (F.R.S.-FNRS) under Grant No. 2.5020.11 and by the Walloon Region. The present research also benefited from computational resources made available on Lucia, the Tier-1 supercomputer of the Walloon Region, infrastructure funded by the Walloon Region under the grant agreement n°1910247. A.P. dedicates this work to N. Palo and M. Palo.

\section{Author Contributions}
H.S.N. conceived the idea and developed the theoretical model. S.C., T.P.V., and C.S. fabricated the structures. A.B. and S.C. carried out the near-field (SNOM) measurements. Z.Y., X.L., and N.R. performed the FDTD simulations. L.B. and A.P. contributed to the optical-bench setup and SLM characterization. A.P., H.S.N., P.N, M.H. and R.H. performed the far-field optical measurements and analyzed the experimental data in comparison with theory. H.S.N., Z.Y., S.C., M.L., and A.P. wrote the initial draft of the manuscript. All authors reviewed, edited, and approved the final version.

\section{Competing interests}
The authors declare no conflicts of interest.

% \section{Conclusion}
% In conclusion, we have demonstrated a novel method for real-space reconfigurable vortex emission from a two-dimensional photonic lattice while maintaining the topology of the momentum space singularity. The lasing action is achieved via the formation of trapped states in the fundamental mode, induced by optical pumping. For a single spot optical pump, we demonstrated the formation of a single trapped state induced lasing, which exhibits a single singularity, both in the real space and the k-space. Furthermore, two-spot pumping was employed to show programmable real-space emissions, which showed dual-mode lasing where the real space exhibited multiple singularities, whereas the momentum space possessed only a single singularity. This experimentally demonstrated that the real space topology can be altered while protecting the momentum space topology.   In addition, the experimental demonstration aligns well with the numerical simulations as well as the analytical model. This type of lasing can be useful for optical tweezing, trapping, and free-space communication. Moreover, due to the formation of trapped states, this work opens up a new perspective for the use of photonic lattices as a quantum emulator for solving the XY Hamiltonian.

\bibliographystyle{apsrev4-2}  % APS-recommended style
\bibliography{references} % Path to your .bib file

\renewcommand{\theequation}{A\arabic{equation}}
\renewcommand{\thefigure}{A\arabic{figure}}
\renewcommand{\thesection}{A\arabic{section}}

\setcounter{equation}{0}
\setcounter{figure}{0}
\setcounter{table}{0}

\newpage
\section{Methods}

\subsection{Analytical model of the magnetic monopolar mode}
\label{Methods:Analytical model}

Following the theoretical framework developed in
Ref.~\cite{nguyen_generalized_2025}, in this section we summarize the effective non-Hermitian model used to
describe the magnetic monopolar mode of the honeycomb PhC slab. Near the $\Gamma$ point, the Bloch field can be expressed as a coherent
superposition of six TE-like guided modes $\ket{n}$ associated with the
first–order reciprocal-lattice vectors ${\vec G}_{n=1..6}$,
\begin{equation}
\vec G_n = \frac{4\pi}{3a}(\cos\phi_n,
\sin\phi_n)^\mathrm{T},
\qquad
\phi_n=\frac{\pi}{3}(1-n),
\end{equation}
each carrying an in-plane polarization vector $\vec p_n=(-\sin\phi_n,\cos\phi_n)^\mathrm{T}$.
All six components correspond to the same fundamental TE guided mode of
the effective homogeneous slab; the lattice simply couples these
harmonics through diffraction and radiations. In this basis eigenmodes of the system is governed by an effective non-Hermitian Hamiltonian, given by:
\begin{widetext}
\begin{equation}
\begin{aligned}
    &\hat{H}_\Gamma(\vec{k})=\omega_\Gamma +\begin{pmatrix}
        v_\Gamma \vert\vec{k}_\parallel\vert\cos{\left(\varphi-\frac{\pi}{3}\right)} &V & W & U & W & V\\
        V&v_\Gamma \vert\vec{k}_\parallel\vert\cos\varphi& V & W & U & W\\
        W & V &v_\Gamma \vert\vec{k}_\parallel\vert\cos{\left(\varphi+\frac{\pi}{3}\right)} &V & W & U\\
        U & W & V&-v_\Gamma \vert\vec{k}_\parallel\vert\cos{\left(\varphi-\frac{\pi}{3}\right)} &V & W\\
        W & U & W & V&-v_\Gamma \vert\vec{k}_\parallel\vert\cos\varphi&V\\
        V & W & U & W & V& -v_\Gamma \vert\vec{k}_\parallel\vert\cos{\left(\varphi+\frac{\pi}{3}\right)}\\
    \end{pmatrix} \\
    &  
    - i\gamma_0\begin{pmatrix}
        1 & \frac{1}{2} & -\frac{1}{2} & -1 & -\frac{1}{2} & \frac{1}{2}\\
        \frac{1}{2} & 1 & \frac{1}{2} & -\frac{1}{2} & -1 & -\frac{1}{2}\\
        -\frac{1}{2} &  \frac{1}{2} & 1 & \frac{1}{2} & -\frac{1}{2} & -1\\
        -1&-\frac{1}{2} &  \frac{1}{2} & 1 & \frac{1}{2} & -\frac{1}{2}\\
        -\frac{1}{2}&-1&-\frac{1}{2} &  \frac{1}{2} & 1 & \frac{1}{2}\\
        \frac{1}{2}&-\frac{1}{2}&-1&-\frac{1}{2} &  \frac{1}{2} & 1  \end{pmatrix}
     \label{eq:H_gamma}
    \end{aligned}
    \end{equation}
\end{widetext}
with $\varphi=\arg(\vec{k}_\parallel)$. Here the coupling strength $U$,$V$ and $W$ represent the diffractive couplings between the guided modes, while $\gamma_0$ represents the radiative coupling of the guided modes to free-space. All of these coupling strength are given by different Fourier components of the periodic dielectric function of the photonic crystal\cite{nguyen_generalized_2025}.

At $\vec{k}_\parallel=\vec{0}$, diagonalizing $\hat H(\vec{0})$ yields six eigenmodes.  
Because the honeycomb geometry satisfies $U<0$, $V<0$, and $W<0$, the eigenvalue
$\omega_0=\omega_\Gamma + U + 2V + 2W$
correspond to the lowest mode  at the $\Gamma$ point.
This mode does not radiate at normal incidence, and therefore, it constitutes a symmetry-protected BIC that forms the parent Bloch resonance for all trapped states studied in this work. 
Its eigenvector is $\mathbf{A}(0)=(1,1,1,1,1,1)^\mathrm{T}$, up to normalization: the six first-order harmonics oscillate in phase.
The other five modes (two dipolar, two quadrupolar, one hexapolar) are
described in detail in Ref.~\cite{nguyen_generalized_2025}.

Expanding the eigenvalue to second order,
\begin{equation}
\omega(\vec{k}_\parallel)
\simeq
\omega_0+
\underbrace{\frac{v_\Gamma^{2}\vert\vec{k}_\parallel\vert^{2}}{2(2U+V+3W)}}_{\alpha\vert\vec{k}_\parallel\vert^{2}}
-i\underbrace{\gamma_0\,\frac{3v_\Gamma^{2}\vert\vec{k}_\parallel\vert^{2}}{2(2U+V+3W)^{2}}}_{\gamma(\vec{k}_\parallel)}
\end{equation}
reveals an isotropic parabolic dispersion.
In out case $2U+V+3W<0$ (see Supplemental Materials), the effective mass
$m=\hbar^2(2U+V+3W)/v_\Gamma^2$ is negative.

To linear order in $\vert\vec{k}_\parallel\vert$, the eigenvector becomes
\begin{equation}
\mathbf{A}(\vec{k}_\parallel)
\simeq
\mathbf{A}(0)
+
\frac{v_\Gamma \vert\vec{k}_\parallel\vert}{\sqrt{2}(2U+V+3W)}
\!\begin{pmatrix}
\sin(\varphi+\tfrac{\pi}{6})\\
\cos\varphi\\
-\sin(\varphi-\tfrac{\pi}{6})\\
-\sin(\varphi+\tfrac{\pi}{6})\\
-\cos\varphi\\
\sin(\varphi-\tfrac{\pi}{6})
\end{pmatrix}\!,
\label{eq:A_small_k}
\end{equation}

The electric field inside the slab (i.e. near field) follows
\begin{equation}
\vec{E}_{\mathrm{near}}(\vec r_\parallel,z)
=
u(z)
\sum_{n=1}^{6}
A_n(\vec{k}_\parallel)\,
e^{i(\vec{G}_n+\vec{k}_\parallel)\!\cdot\vec r_\parallel}\,
\vec p_n ,
\label{eq:NF_monopole}
\end{equation}
where $u(z)$ is the vertically guided mode profile.
For the monopolar BIC at $\Gamma$, $A_n(0)=1$ for all $n$, giving
\begin{equation}
\vec{E}_{\mathrm{near}}(\vec r_\parallel,z)
\propto
u(z)\sum_{n=1}^{6}e^{i\vec{G}_n\cdot\vec r_\parallel}\vec p_n .
\end{equation}

The magnetic field obtained from Maxwell’s equations,
\begin{equation}
H_z(\vec r_\parallel,z)\propto
u(z)\sum_{n=1}^{6}e^{i\vec{G}_n\cdot\vec r_\parallel} ,
\end{equation}
produces a single-lobed, same-sign $H_z$ distribution in each
hexagonal unit cell, while the in-plane electric field $\vec{E}_{\mathrm{near}}(\vec r_\parallel,z)$ forms a circulating pattern around the cell center.  
This fully symmetric $A_1$ texture — single-lobed $H_z$ with circulating
$\vec{E}_\parallel$ is the origin of the term "magnetic monopolar".

The radiated field is the guided-mode
superposition projected onto the outgoing channel via radiative coupling:
\begin{equation}
\vec{E}_{\mathrm{far}}(\vec{k}_\parallel)
\propto
\sqrt{\gamma_0}\Bigl[\sum_n A_n(\vec{k}_\parallel)\vec p_n\Bigr].
\end{equation}
Using Eq.~\eqref{eq:A_small_k} one obtains, to lowest order in $\vert\vec{k}_\parallel\vert$,
\begin{equation}
\vec{E}_{\mathrm{far}}(\vec{k}_\parallel)
\propto
\vert\vec{k}_\parallel\vert
\begin{pmatrix}
-\sin \varphi\\
\cos \varphi
\end{pmatrix}
=
\vert\vec{k}_\parallel\vert\,\hat{\boldsymbol u}_\varphi,
\label{eq: Methods analytical}
\end{equation}
and therefore the monopolar mode is a  BIC of
topological charge $+1$ at $\Gamma$.  
Moreover, the radiative loss rate  $\gamma(\vec{k}_\parallel)$ can also be evaluated from the emitted intensity
\begin{equation}
    \vert\vec{E}_{\mathrm{far}}(\vec{k}_\parallel)\vert^2
\propto \gamma_0\,\frac{3v_\Gamma^{2}\vert\vec{k}_\parallel\vert^{2}}{2(2U+V+3W)^{2}}.
\end{equation}. This provides the same expression of  $\gamma(\vec{k}_\parallel)$ as the one obtained from the eigenvalue. 

\subsection{Far-field in real space of the trapped state}

 For a trapped state of envelope
$F(\vec r_\parallel)$ in real space, whose Fourier transform is
$F(\vec k_\parallel)$, the farfield in momentum space is given by $    \vec{E}^{\text{trap}}_{\text{far}}(\vec{k}_\parallel)
\propto
\vec{E}_{\mathrm{far}}(\vec{k}_\parallel)\,F(\vert\vec{k}_\parallel\vert)$. Moreover, from Eq.~\eqref{eq: Methods analytical}, the radiation of the monopolar BIC near
$\Gamma$ scales as
$\vec{E}_{\mathrm{far}}(\vec{k}_\parallel)
\propto
\vert\vec{k}_\parallel\vert\,\hat{\boldsymbol u}_\varphi=\begin{pmatrix}
-k_y\\
k_x
\end{pmatrix}$, leading to
\begin{equation}
  \vec{E}^{\text{trap}}_{\text{far}}(\vec{k}_\parallel)
\propto
F(\vert\vec{k}_\parallel\vert)\begin{pmatrix}
-k_y\\
k_x
\end{pmatrix}.
\end{equation}

The real-space far-field distribution is obtained by inverse Fourier transform,
\begin{equation}
\vec{E}^{\text{trap}}_{\text{far}}(\vec{r}_\parallel)
=
\int
F(\vert\vec{k}_\parallel\vert)\begin{pmatrix}
-k_y\\
k_x
\end{pmatrix}
e^{i\vec{k}_\parallel\cdot\vec{r}_\parallel}
\,dk_x\,dk_y .
\end{equation}

Using the Fourier derivative identities $k_x e^{i\vec{k}_\parallel\cdot\vec{r}_\parallel}
=
-i\,\partial_x e^{i\vec{k}_\parallel\cdot\vec{r}_\parallel}$ and $
k_y e^{i\vec{k}_\parallel\cdot\vec{r}_\parallel}
=
-i\,\partial_y e^{i\vec{k}_\parallel\cdot\vec{r}_\parallel}$, one obtains $\vec{E}_{\mathrm{far}}(\vec{r}_\parallel)
\propto
\begin{pmatrix}
-\partial_y F(\vec{r}_\parallel)\\
\partial_x F(\vec{r}_\parallel)
\end{pmatrix}$. This can be written compactly as
\begin{equation}
\vec{E}^{\text{trap}}_{\text{far}}(\vec{r}_\parallel)
\propto
\hat{\mathbf z}\times\nabla F(\vec{r}_\parallel).
\end{equation}

If the trapped-state envelope is radially symmetric, $F(\vec r_\parallel)=F(\vert\vec r_\parallel\vert)$, then $\nabla F
= F'(\vert\vec r_\parallel\vert)\,\hat{\boldsymbol u}_r$. This yields an even simpler expression, given by $\vec{E}_{\mathrm{far}}(\vec{r}_\parallel)
\propto
F'(\vert\vec r_\parallel\vert)\,\hat{\boldsymbol u}_\varphi$.

Therefore the real-space far-field corresponds to the azimuthal rotation of the
gradient of the envelope function. For any radially symmetric trapped state,
the emission vanishes at the center and forms a ring-shaped intensity profile
determined by the radial derivative of the envelope.

\subsection{Determination of the potential amplitude $V_0$}
\label{Methods: Potential V0}

The envelope function and the corresponding eigenenergy of the trapped state are obtained by numerically solving the two-dimensional Schr\"odinger equation $-\frac{\hbar^2}{2m}\,\nabla_{\vec{r}_\parallel}^2 F(\vec{r}_\parallel) + V(\vec{r}_\parallel)\,F(\vec{r}_\parallel) = \Delta E\,F(\vec{r}_\parallel)$, where $F(\vec{r}_\parallel)$ denotes the envelope function, $V(\vec{r}_\parallel)$ the effective potential induced by the pump, and $E$ the corresponding eigen-energy.

For single-spot pumping, the carrier-induced potential is modeled as a Gaussian profile $V(\vec r_\parallel)=V_0 \exp\!\left(-\frac{|\vec r_\parallel|^2}{\sigma^2}\right)$, characterized by the pump waist $\sigma$ and the potential amplitude $V_0$. The waist $\sigma$ is determined experimentally from a Gaussian fit to the real-space pump-intensity profile on the sample. The potential amplitude $V_0$ is then treated as a fitting parameter and adjusted until the eigenenergy obtained from the Schr\"odinger solver reproduces the confinement energy $\Delta E$ measured experimentally in the saturated regime [Fig.~\ref{fig:3}(d)].

\subsection{Experimental setup for farfield measurements}
\label{Methods: setup}

Energy–momentum dispersions were measured using a custom-built back-focal-plane (BFP) imaging setup.  
The Fourier plane of a $\times 100$ microscope objective (NA = 0.8) was imaged onto the entrance slit of a grating spectrometer (HRS-500MS-NI, Princeton Instruments).  
The sample was mounted such that its $x$-axis, corresponding to the $\Gamma$–$K$ direction of the honeycomb lattice, was aligned with the slit.  
The spectrum dispersed by the grating was detected using a two-dimensional InGaAs camera (NIRvana HS, Princeton Instruments), providing simultaneous access to $k_x$ and optical energy.  
To measure the $\Gamma$–$M$ dispersion, the sample was rotated such that its $y$-axis aligned with the slit, providing access to $k_y$. Angle-resolved reflectivity [Fig.~\ref{fig:2}(d)] and angle-resolved photoluminescence [Figs.~\ref{fig:3}(b) and \ref{fig:5}(b)] are acquired using the same setup but with different illumination sources: a broadband halogen lamp for reflectivity and a femtosecond laser (800nm, Coherent Chameleon Ultra II) for photoluminescence.

\begin{figure*}
    \centering
    \includegraphics[width=1\linewidth]{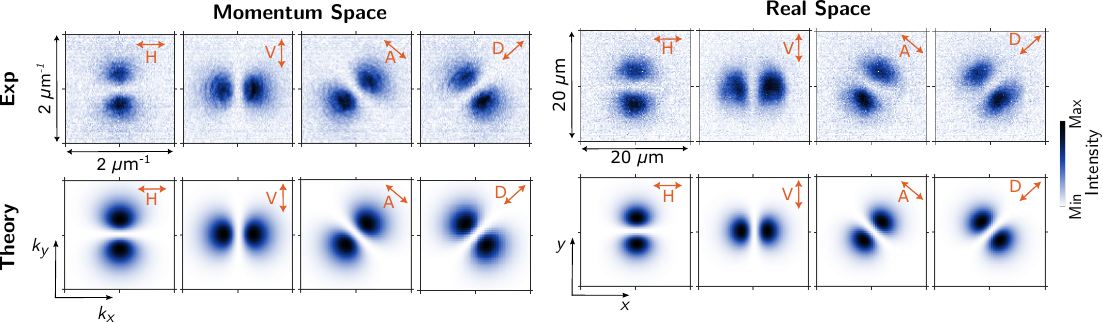}
   
       \caption{\textbf{Polarization-resolved maps under single-spot pumping.}
Experimental and theoretical polarization-resolved far-field lasing images in both real and momentum space for a pump waist $\sigma = 3.2~\mu$m. A linear polarizer is placed in front of the InGaAs camera (see Methods) and rotated to 90$^\circ$ (H \rotatebox{0}{$\leftrightarrow$} ), 0$^\circ$ (V \rotatebox{90}{$\leftrightarrow$}), -45$^\circ$ (A \rotatebox{135}{$\leftrightarrow$}), and 45$^\circ$ (D \rotatebox{45}{$\leftrightarrow$}). The corresponding polarization-integrated images without a polarizer are shown in Fig.~\ref{fig:3}(e–h).}
    \label{fig:Extended_Figure_1}
\end{figure*}

Real-space and momentum-space far-field intensity and polarization maps  
[Figs.~\ref{fig:2}(e-g), \ref{fig:4}(d-g), and \ref{fig:Extended_Figure_1}] are obtained on the same optical platform operated in a tomographic configuration, in which either the BFP plane or the object plane was scanned relative to the spectrometer slit while selecting different polarization analyzers.  
This method, which yields quantitative polarization-resolved images in both $(k_x,k_y)$ and $(x,y)$, is described in detail in our previous study~\cite{cueff_fourier_2024}. For a detailed description of the optical setup refer to the Supplementary Information.

\subsection{Polarization orientation}

The polarization azimuth $\phi$ was extracted from polarization-resolved intensity measurements in the horizontal (H), vertical (V), diagonal (D), and antidiagonal (A) bases. From these measurements, the normalized linear Stokes parameters were computed as
\begin{equation}
S_1 = \frac{I_H - I_V}{I_H + I_V}, \qquad
S_2 = \frac{I_D - I_A}{I_D + I_A}.
\end{equation}

The degree of polarization $p$ is given by
\begin{equation}
p = \sqrt{S_1^2 + S_2^2}.
\end{equation}

The polarization azimuth/orientation $\phi$ was evaluated as
\begin{equation}
\phi = \frac{1}{2}\,\mathrm{sign}(S_2)\,
\arccos\!\left(\frac{S_1}{p}\right),
\end{equation}
and expressed in degrees. Points where $\phi$ is undefined are called polarization singularities.

The topological charge $q$ of these singularities was determined from the winding of the polarization azimuth around a closed contour $\mathcal{C}$ enclosing the singularity,
\begin{equation}
q = \frac{1}{2\pi} \oint_{\mathcal{C}} \nabla \phi \cdot d\mathbf{l}.
\end{equation}
This quantity measures the net rotation of $\phi$ along the contour.

\newpage

\begin{figure*}
    \centering
    \includegraphics[width=1\linewidth]{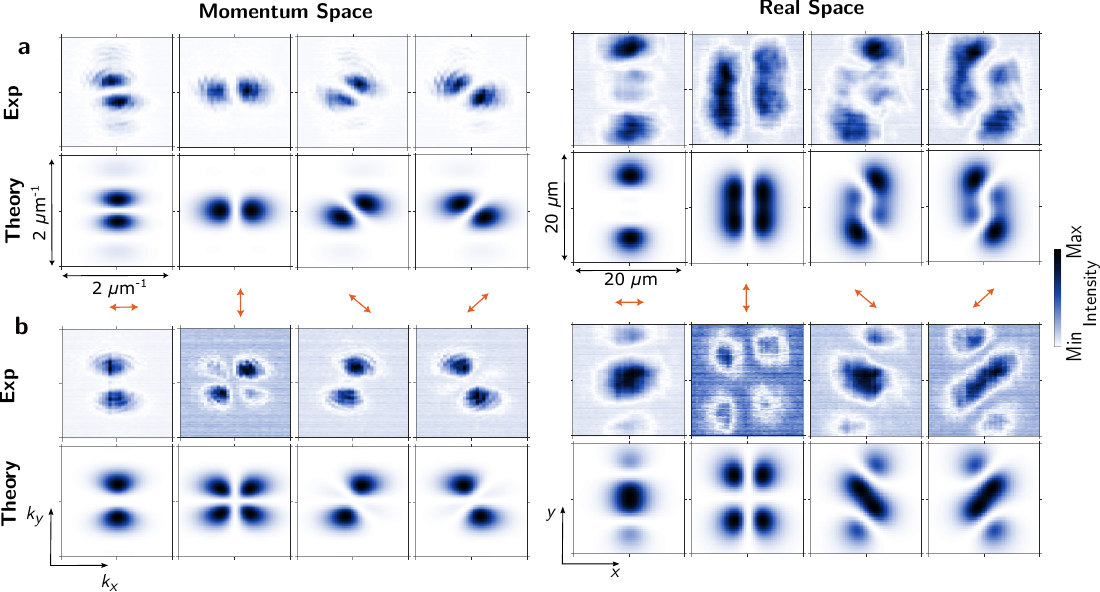}
\caption{\textbf{Polarization maps for double-spot pumping.}
Polarization-resolved far-field lasing images for the \textbf{(a)} bonding and \textbf{(b)} antibonding modes in real and momentum space, corresponding to the data shown in Fig.~\ref{fig:5}(d–g). A linear polarizer placed in front of the detector is rotated to 90$^\circ$ (H \rotatebox{0}{$\leftrightarrow$} ), 0$^\circ$ (V \rotatebox{90}{$\leftrightarrow$}), -45$^\circ$ (A \rotatebox{135}{$\leftrightarrow$}), and 45$^\circ$ (D \rotatebox{45}{$\leftrightarrow$}). The top and bottom rows display the experimental and analytical results, respectively.}
    \label{fig:Extended_Figure_2}
\end{figure*}

\begin{figure*}
    \centering
    \includegraphics[width=0.6\linewidth]{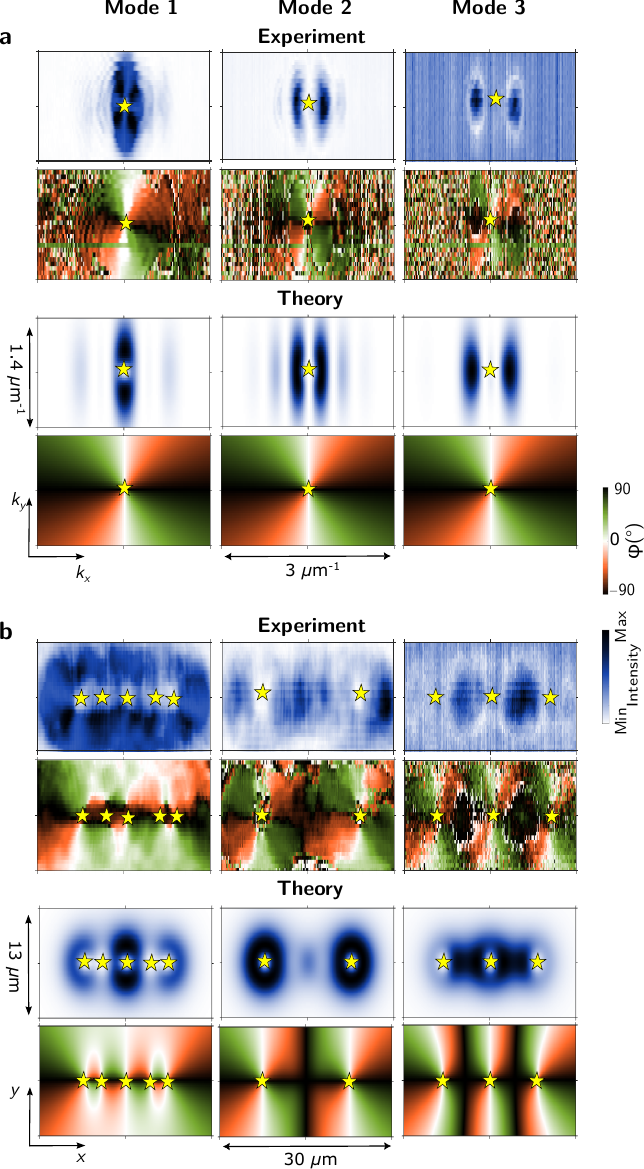}
     \caption{\textbf{Three-spot pumping and multi-singularity trapped modes.}
    Experimental and theoretical far-field characterization of the three trapped
    modes formed under three-spot pumping with spot separation $L=7.5~\mu\mathrm{m}$, spot size $\sigma$ = 2.5 $\mu$m and V$_0$= 5.5 meV.
    Each column corresponds to one of the three trapped eigenstates supported by the
    triple-well potential. For each mode, the panels show (from top to bottom):
    \textbf{(a)} experimental momentum-space intensity, experimental momentum-space polarization texture,
    theoretical momentum-space intensity, theoretical momentum-space polarization texture.
    The same sequence is used for the \textbf{(b)} real-space emission. As expected, all
    modes preserve the same momentum-space vortex at $\Gamma$, inherited from the
    monopolar BIC of the underlying Bloch resonance. In contrast, the number and spatial
  arrangement of real-space polarization singularities (yellow stars) directly
    follow from the nodal structure of the pump-induced envelope function: five
    singularities for the ground state (Mode~1), two for the first excited state
    (Mode~2), and three for the second excited state (Mode~3).  }
    \label{fig:Extended_Figure_3}
\end{figure*}

\begin{figure*}
    \centering
    \includegraphics[width=1.0\linewidth]{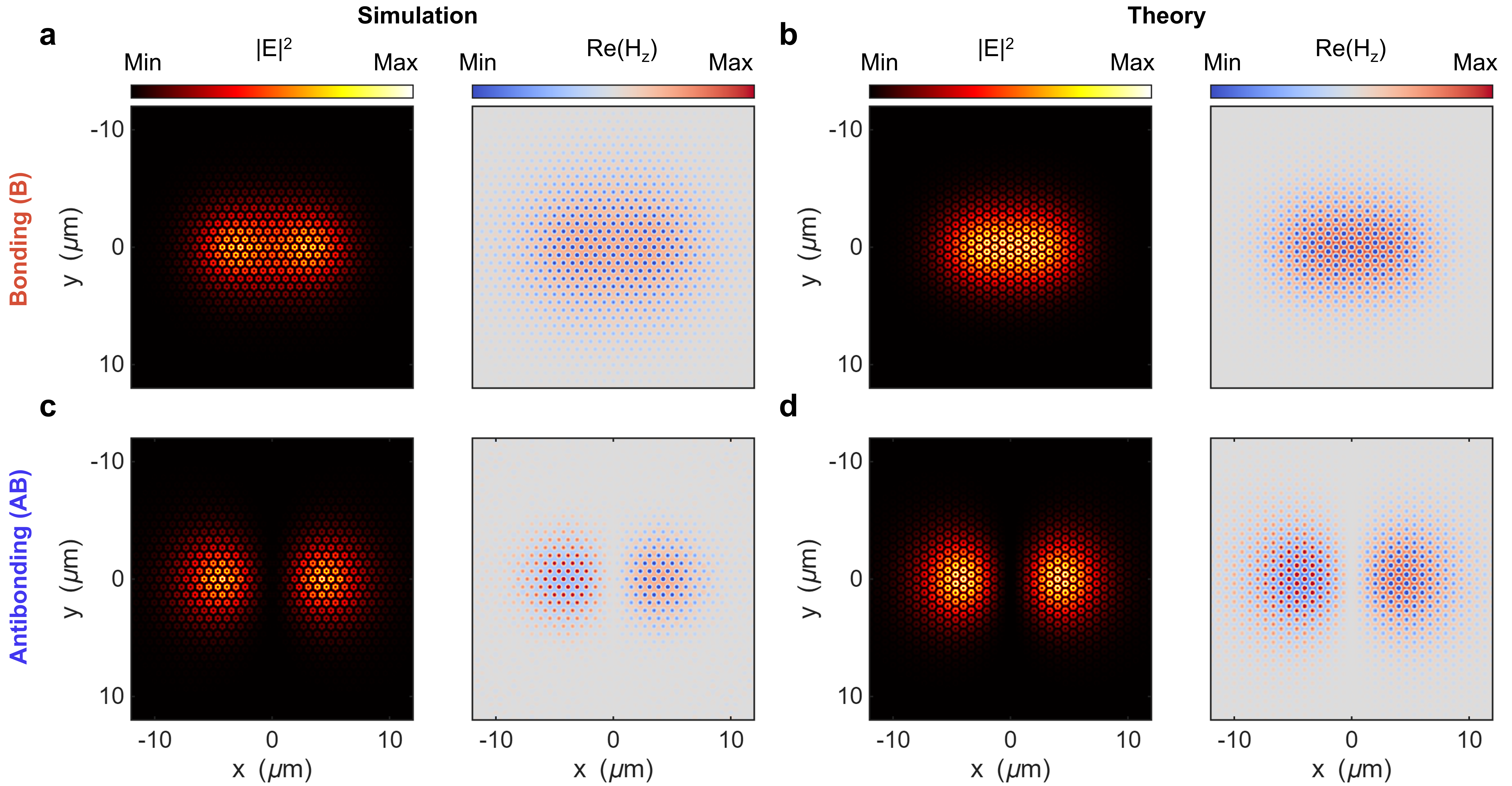}
    \caption{Near-field map of the trapped states under
the coupled Gaussian pumps $V(x,y)=V_0\!\left[e^{-((x-L/2)^2-y^2)/\sigma^2}+e^{-((x+L/2)^2+y^2)/\sigma^2}\right]$. The symmetric and antisymmetric against $x=0$ exhibit the feature of bonding and antibonding mode.}
    \label{fig:EF-Near-Field-couplemode}
\end{figure*}

\begin{figure*}
    \centering
    \includegraphics[width=1.0\linewidth]{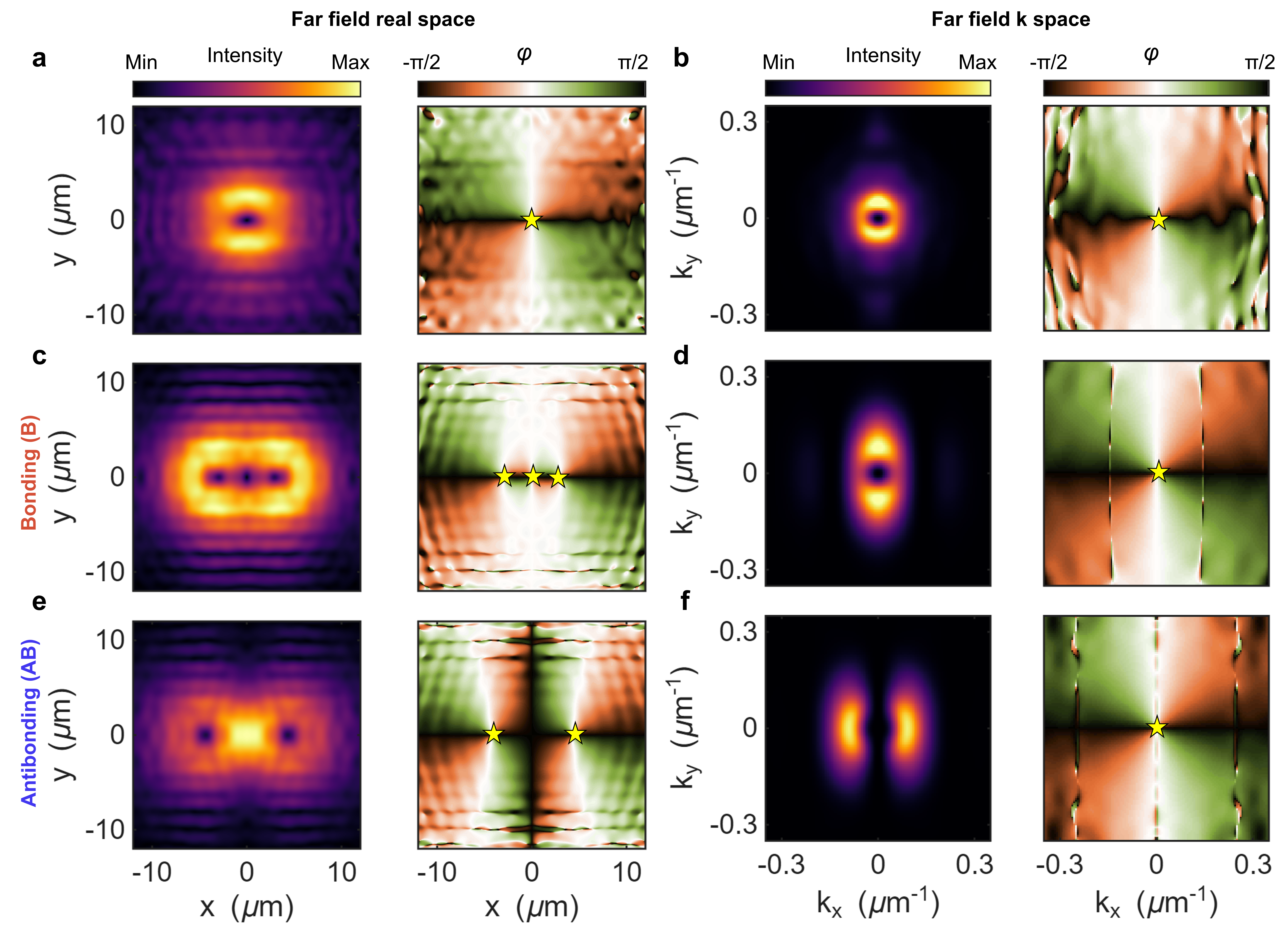}
    \caption{Numerical simulation results of far-field patterns and polarization angles in real and momentum space. \textbf{(a-b)} Single barrier with $\sigma= 3.2$µm. \textbf{(c-f)} Bonding and Antibonding modes from double barriers with $\sigma= 4$µm and $L=$8µm.}
    \label{fig:EF-Sim-Vortex}
\end{figure*}

%\begin{figure*}
%    \centering
%    \includegraphics[width=1.0\linewidth]{SM-Figures/Sim-Vortex.png}
%    \caption{Numerical simulation results of far-field patterns and polarization angles in real and momentum space. \textbf{(a-b)} Single barrier with $\sigma= 3.2$µm. \textbf{(c-f)} Bonding and Antibonding modes from double barriers with $\sigma= 4$µm and $L=$8µm.}
%    \label{fig:ED-vortex}
%\end{figure*}

\end{document}